\documentclass[12pt]{article}
\usepackage{amssymb}

\def\be{\begin{equation}}
\def\ee{\end{equation}}
\def\bea{\begin{eqnarray}}
\def\eea{\end{eqnarray}}
\def\ba{\begin{array}}
\def\ea{\end{array}}

\def\o{\omega}

\def\p{\partial}

\def\p{\partial}

\def\np#1{{\sl Nucl.~Phys.~\bf B#1}}
\def\pl#1{{\sl Phys.~Lett.~\bf B#1}}
\def\pr#1{{\sl Phys.~Rev.~\bf D#1}}

\def\cqg#1{{\sl Class.~Quant.~Grav.~\bf #1}}

\catcode`\@=11
\def\@citex[#1]#2{%
\if@filesw \immediate \write \@auxout {\string \citation {#2}}\fi
\@tempcntb\m@ne \let\@h@ld\relax \def\@citea{}%
\@cite{%
  \@for \@citeb:=#2\do {%
    \@ifundefined {b@\@citeb}%
      {\@h@ld\@citea\@tempcntb\m@ne{\bf ?}%
      \@warning {Citation `\@citeb ' on page \thepage \space undefined}}%
      {\@tempcnta\@tempcntb \advance\@tempcnta\@ne%
      \@tempcntb\number\csname b@\@citeb \endcsname \relax%
      \ifnum\@tempcnta=\@tempcntb 
        \ifx\@h@ld\relax%
          \edef \@h@ld{\@citea\csname b@\@citeb\endcsname}%
        \else%
          \edef\@h@ld{\ifmmode{-}\else--\fi\csname b@\@citeb\endcsname}%
        \fi%
      \else
        \@h@ld\@citea\csname b@\@citeb \endcsname%
        \let\@h@ld\relax%
      \fi}%
    \def\@citea{,\penalty\@highpenalty\,}%
  }\@h@ld
}{#1}}

\def\@citeb#1#2{{[#1]\if@tempswa , #2\fi}}
\def\@citeu#1#2{{$^{#1}$\if@tempswa , #2\fi }}
\def\@citep#1#2{{#1\if@tempswa , #2\fi}}

\def\bcites{         
        \catcode`\@=11
        \let\@cite=\@citeb
        \catcode`\@=12
}

\def\upcites{         
        \catcode`\@=11
        \let\@cite=\@citeu
        \catcode`\@=12
}

\def\plaincites{      
        \catcode`\@=11
        \let\@cite=\@citep
        \catcode`\@=12
}

\newcount\hour
\newcount\minute
\newtoks\amorpm
\hour=\time\divide\hour by 60
\minute=\time{\multiply\hour by 60 \global\advance\minute by-\hour}
\edef\standardtime{{\ifnum\hour<12 \global\amorpm={am}%
        \else\global\amorpm={pm}\advance\hour by-12 \fi
        \ifnum\hour=0 \hour=12 \fi
        \number\hour:\ifnum\minute<10 0\fi\number\minute\the\amorpm}}
\edef\militarytime{\number\hour:\ifnum\minute<10 0\fi\number\minute}

\def\draftlabel#1{{\@bsphack\if@filesw {\let\thepage\relax
   \xdef\@gtempa{\write\@auxout{\string
      \newlabel{#1}{{\@currentlabel}{\thepage}}}}}\@gtempa
   \if@nobreak \ifvmode\nobreak\fi\fi\fi\@esphack}
        \gdef\@eqnlabel{#1}}
\def\@eqnlabel{}
\def\@vacuum{}
\def\marginnote#1{}
\def\draftmarginnote#1{\marginpar{\raggedright\scriptsize\tt#1}}
\def\draft{
        \pagestyle{plain}
        \overfullrule=2pt
        \oddsidemargin -.5truein
        \def\@oddhead{\sl \phantom{\today\quad\militarytime} \hfil
        \smash{\Large\sl DRAFT} \hfil \today\quad\militarytime}
        \let\@evenhead\@oddhead
        \let\label=\draftlabel
        \let\marginnote=\draftmarginnote
        \def\ps@empty{\let\@mkboth\@gobbletwo
        \def\@oddfoot{\hfil \smash{\Large\sl DRAFT} \hfil}
        \let\@evenfoot\@oddhead}
        \def\@eqnnum{(\theequation)\rlap{\kern\marginparsep\tt\@eqnlabel}%
        \global\let\@eqnlabel\@vacuum}  }

\title{Poincar\'e recurrences of Schwarzschild black holes\footnote{Research supported in part by the DoE under grant DE-FG05-91ER40627.}}
\author{George Siopsis\footnote{siopsis@tennessee.edu}\\
\em Department of Physics
and Astronomy, \\
\em The University of Tennessee, Knoxville, \\
\em TN 37996 - 1200, USA.
}
\date{May 2007}
\begin{document}

\maketitle
\vspace{-3.5in}\hfill UTHET-07-0201\vspace{3.5in}

\abstract{We discuss massive scalar perturbations of a Schwarzschild black hole.
We argue that quantum effects alter the effective potential near the horizon
resulting in Poincar\'e recurrences in Green functions.
Results at the semi-classical level are independent of the details of the modification of the potential provided its minimum near the horizon is inversely proportional to the square of the
Poincar\'e time.
This modification may be viewed as a change in the near-horizon geometry.
We consider explicitly the examples of a brick wall, a smooth cutoff and a wormhole-like modification showing that they all lead to the same results at leading order.
}
\newpage
\section{Introduction}
\label{sec1}

The decay of a black hole through Hawking radiation~\cite{bibH} has led to the
information loss paradox questioning the unitary evolution of a black hole.
This issue has attracted a lot of attention and despite considerable effort~\cite{bibbh1,bibbh2,bibbh3} it remains unsettled.
In principle, its resolution lies within string theory which is a unitary quantum theory containing gravity.
Unfortunately, calculations beyond the semi-classical order are cumbersome.
Moreover, it appears that perturbation theory is not adequate for understanding the information loss paradox~\cite{bibego,bibsol2}.
The argument is based on the Poincar\'e recurrence theorem.
Since the entropy of the black hole is finite, one expects that, once perturbed, the black hole will never relax back to its original state if its evolution is unitary.
The perturbation should be quasi-periodic with a large period
\be\label{eqtP} t_P \sim e^S \ee
where $S$ is the entropy of the black hole.
For times
$t\ll t_P$, the system will look like it is decaying back to thermal equilibrium,
which explains the blackbody spectrum of Hawking radiation in the semi-classical regime.
However, if one waits long enough,
for $t\gtrsim t_P$, the perturbation should return to its original state (or close) an
infinite number of times.
This behavior should be evident in any correlator.
The thermal correlator one obtains semi-classically contains $\mathcal{O} (1/t_P)$ corrections which become important only for times $t\gtrsim t_P$~\cite{bibq13a}.
Given the universal Bekenstein-Hawking form of the entropy
\be\label{eqE} S_{BH} = \frac{\mathcal{A}_h}{4G} \ee
where $\mathcal{A}_h$ is the area of the horizon and $G$ is Newton's constant,
it is evident that the Poincar\'e recurrence time (\ref{eqtP}) is generically
\be t_P \sim \mathcal{O} (e^{1/G}) \ee
It follows that the corrections which are responsible for restoring unitarity to the thermal semi-classical state are $\mathcal{O} (e^{-1/G})$.
Their understanding requires inclusion of non-perturbative effects.
Studying these late-time effects is important not only for showing that quantum gravity is a unitary theory, but also in view of the possibility of observing decaying black holes at the LHC.

The finiteness the Poincar\'e recurrence time relies on the finiteness of the entropy. The latter is complicated by the introduction of a perturbation.
Once a matter field is added, it contributes to the entropy an infinite quantity
due to the
infinite blue-shift experienced by an in-falling object near the horizon.
To tackle the  infinities, 't Hooft introduced an artificial ``brick wall'' just outside the horizon beyond which a particle cannot propagate~\cite{bibtH}.
It was subsequently understood that infinities may be absorbed by the gravitational constants
and the total entropy is finite when expressed in terms of physical parameters as in
any renormalizable field theory~\cite{bibE1,bibE2}.
The form~(\ref{eqE}) of the entropy, including these quantum effects, remains unchanged.

In the case of asymptotically AdS space-times,
the AdS/CFT correspondence~\cite{bibadscft} offers an additional tool in the study of unitarity,
because the CFT on the boundary of AdS is a unitary field theory~\cite{bibq13a,bibpr1,bibpr2,bibpr3,bibpr4,bibpr5,bibsol}.
It was argued by Solodukhin~\cite{bibsol} that quantum effects replace the horizon by a wormhole of narrow throat $\sim \mathcal{O} (1/t_P)$.

There is no similar correspondence principle in asymptotically flat space-times.
However, the Poincar\'e recurrence theorem should still hold. To lend support
to this claim, we calculated the two-point function of a massive scalar
field in the background of a two-dimensional dilaton black hole \cite{bibego2d}.
By concentrating on the dynamics near the horizon, we argued that the
effective potential was modified by quantum effects.
For an explicit calculation, we concentrated on a wormhole modification. However,
the results were independent of the detailed shape of the effective potential.
We demonstrated that the two-point function exhibited Poincar\'e recurrences, as
expected.

Here we extend the discussion of \cite{bibego2d} to the physically more relevant case of a four-dimensional Schwarzschild black hole.
We consider a massive scalar field and modify its effective potential near the horizon. We calculate the propagator using the modified potential semi-classically and arrive at expressions exhibiting Poincar\'e recurrences.
Our results are independent of the details of the modification at leading order. Thus, we provide evidence that Poincar\'e recurrences are a generic feature of such modifications.

In detail, we start in section \ref{sec2} with a review of the effective action of the matter field and its effect on the entropy of the black hole following~\cite{bibE2}.
We then calculate the two-point function and obtain the decaying behavior characteristic of a system in thermal equilibrium.
In section \ref{sec3} we discuss various modifications of the effective potential near the horizon and show that they all lead to the same results for the two-point function
including Poincar\'e recurrences.
Finally in section \ref{sec4} we discuss the generic features of our argument and present our conclusions.

\section{Massive scalar field}
\label{sec2}

In this section we consider a gravitational field coupled to a massive scalar field
and show how the latter renormalizes the gravitational constants parametrizing
the gravitational action.
We also calculate the effcts on thermodynamic quantities in the semi-classical approximation.
This is a review of the discussion in \cite{bibE2} and is presented for completeness and to fix the notation.
We then calculate the two-point function of the matter field and exhibit its decaying behavior in the thermal bath.

Consider the Einstein-Hilbert gravitational action in four dimensions,
\be\label{eqEH} I_{gr} = \frac{1}{16\pi G} \int d^4 x \sqrt{-g} \ (R-2\Lambda) \ee
where $G$ is Newton's constant and $\Lambda$ is the cosmological constant.
Let us add a massive scalar field $\phi$ of mass $m$ with action
\be\label{eqSm} I_{matter} = \frac{1}{2} \int d^4 x \sqrt{-g} \ [ (\nabla\phi)^2 - m^2 \phi^2 ] \ee
After integrating over the scalar field in the path integral, we arrive at an
effective action which is divergent.
The divergences may be eliminated by a Pauli-Villars regularization~\cite{bibE2}.
We need to add a pair of scalar fields of (large) mass $M_1$, a pair of scalar fields of mass $M_2$ each obeying
wrong statistics and a fifth scalar field of wrong statistics and mass $M_3$, where
\be\label{eqMMM} M_1 = \sqrt{3 M^2+m^2}\ \ , \ \ \ \
M_2 = \sqrt{M^2+m^2} \ \ , \ \ \ \ M_3 = \sqrt{4M^2+m^2} \ee
We obtain the effective action
\be\label{eqIphi} W_{matter} = \frac{1}{32\pi^2}\int d^4 x \sqrt{-g} (A_0a_0(x) + A_1a_1(x)+A_2a_2(x) +\dots) \ee
where the dots represent finite contributions.
The various invariants are
\be\label{eqaaa} a_0 (x)=1 \ \ , \ \ \ \ a_1(x) = \frac{1}{6} R\ \ , \ \ \ \
 a_2(x) = \frac{1}{180} R^{abcd} R_{abcd} - \frac{1}{180} R^{ab} R_{ab} + \frac{1}{30} \Box R + \frac{1}{72} R^2 \ee
and their coefficients are given, respectively, by
\bea\label{eqAAA} A_0 &=& 2M^4 \ln \frac{M_2M_3^8}{M_1^9} + 4m^2M^2 \ln \frac{M_2M_3^2}{M_1^3} + m^4 \ln \frac{M_2^2M_3}{mM_1^2}\nonumber\\
A_1 &=& 4M^2 \ln \frac{M_1^3}{M_2M_3^2} +m^2 \ln \frac{mM_1^2}{M_2^2M_3}\nonumber\\
A_2 &=& 2\ln \frac{M_3M_2^2}{mM_1^2} \eea
in terms of the mass parameters (\ref{eqMMM}). They diverge as $M\to\infty$.

The first two terms in the effective action (\ref{eqIphi}) are of the same type as the terms in the Einstein-Hilbart action (\ref{eqEH}) and lead to a renormalization of Newton's constant $G$,
\be\label{eqGren} \frac{1}{G_R} = \frac{1}{G} + \frac{1}{12\pi} A_1 \ee
and similarly for the cosmological constant $\Lambda$.
The third term leads to a renormalization of additional terms in the gravitational effective action,
\be\label{eqEHn} \delta I_{gr} = \frac{1}{4\pi} \int d^4 x\sqrt{-g} \left[ \kappa_1 R^2 + \kappa_2 R_{ab} R^{ab} + \kappa_3 R_{abcd} R^{abcd} \right] \ee
Notice that there is no term proportional to $\Box R$, because it is a total derivative.
We deduce from (\ref{eqaaa})
\be\label{eqk} \kappa_{1R} = \kappa_1 + \frac{1}{576\pi} A_2 \ \ , \ \ \ \
\kappa_{2R} = \kappa_2 - \frac{1}{1440\pi} A_2 \ \ , \ \ \ \
\kappa_{3R} = \kappa_3 + \frac{1}{1440\pi} A_2 \ee
The Schwarzschild black hole solves the field equations of the Einstein-Hilbert action (\ref{eqEH}) with vanishing cosmological constant. The metric is
\be\label{eqmetric} ds^2 = -f(r) dt^2 + \frac{dr^2}{f(r)} + r^2 d\Omega_2^2
\ \ , \ \ \ \ f(r) = 1 - \frac{r_0}{r} \ee
The horizon is located at $r=r_0$.
The Hawking temperature is given by
\be\label{eqTH} T_H = \frac{f'(r_0)}{4\pi} = \frac{1}{4\pi r_0}\ee
The entropy of the gravitational system has a contribution of the Bekenstein-Hawking
form (\ref{eqE})
where $\mathcal{A}_h = 4\pi r_0^2$.
In what follows, we shall choose units so that the radius of the horizon is $r_0 =1$.

The entropy receives an additional contribution (independent of the area of the horizon) from the addtional terms (\ref{eqEHn}) in the gravitational effective action,
\be\label{eqE2} \delta S_{gr} = -8\pi \kappa_2 +16\pi \kappa_3 \ee
Notice that the first term which is proportional to $\kappa_1$ does not contribute because $R=0$ for the black hole (\ref{eqmetric}).

The above expressions are modified by the addition of the matter field with dynamics determined by the action (\ref{eqSm}).
By varying the action~(\ref{eqSm}), we obtain the wave equation for a massive scalar field in the background~(\ref{eqmetric}),
\be\label{eqwt0} \Box \phi + m^2 \phi = 0 \ee
After decomposing the scalar field,
\be\label{eqscf} \phi (r,\Omega_2, t) = \frac{1}{r} \Phi_{\omega l} (r) Y_{lm} (\Omega_2) e^{-i\o t} \ee
the wave equation (\ref{eqwt0}) can be written in a Schr\"odinger-like form
\be\label{eqwt} - \Phi_{\omega l}'' + (V[r(z)] - \o^2) \Phi_{\omega l} = 0 \ee 
where a prime denotes differentiation with respect to the ``tortoise coordinate'' $z$ given by
\be\label{eqtor} z = \int \frac{dr}{f(r)} = r-1 + \ln ( r -1 ) \ee
and the potential is
\be\label{eq20} V(r) = f(r) \left[ \frac{l(l+1)}{r^2} + \frac{1}{r^3} + m^2 \right] \ee
The
horizon is located at $z= - \infty$. To study the behavior of the scalar field in the observable region outside the horizon, we need to solve the Schr\"odinger equation (\ref{eqwt}) along the entire real axis.
For a complete understanding of the quantum system, we also need to study a similar equation inside the horizon and then match the wavefunctions across the horizon. We shall not address this issue here.

In the WKB approximation, the wave-function for $z<z_0$, where $V[r(z_0)] = \o^2$, is
\be\label{eq24r} \Phi_{\omega l}^{\mathrm{WKB}} (z) = \mathcal{C} \sin \left( p_m(z) + \frac{\pi}{4} \right)\ \ , \ \ \ \ p_m(z) = \int_z^{z_0} dz'\ \sqrt{\o^2 -
V[r(z')] }\ee
where $\mathcal{C}$ is a normalization constant.
We have
\be\label{eq25r} p_m(z) = -\omega z + \theta_m(\o) + \dots \ \ , \ \ \ \
\theta_m (\o) = \o z_0 + \int_{-\infty}^{z_0} dz'\ \left( \sqrt{\o^2 -
V[r(z')] } - \o \right) \ee
where the dots represent terms that vanish as we approach the horizon.
The wave-function (\ref{eq24r}) near the horizon becomes
\be\label{eqwzWr} \Phi_{\omega l}^{\mathrm{WKB}} (z) \approx \mathcal{C}' \left[ e^{i\omega z} + e^{i\left( 2\theta_m(\omega) - \frac{\pi}{2} \right)} e^{-i\omega z} \right] \ee
where $\mathcal{C}' = \frac{1}{2} e^{-i(\theta_m(\omega) - \frac{\pi}{4})} \mathcal{C}$.
Thus near the horizon we obtain plane waves of wavenumber
\be\label{eqdis0} k = \omega \ee
This is the dispersion relation for a {\em massless} free field. The mass $m$ of the scalar field does not contribute in this region because of the infinite blue shift.
We shall argue in section \ref{sec3} that quantum effects alter this dispersion relation by an effective small mass term.
The latter corresponds to a large but finite blue shift at the horizon and is inversely proportional to the Poincar\'e time.
It is therefore much smaller than the mass $m$, yet finite.

The phase $\theta_m(\omega)$ may be calculated by expanding in $\omega$ which amounts to an expansion of the potential near the horizon. We obtain
\be\label{eqtheta} \theta_m (\omega) = -2\omega +\o \ln \frac{4\o^2}{\mathbf{v}_m} + \frac{2(2\mathbf{v}_m +1)}{3\mathbf{v}_m^2}\ \o^3 + \dots \ \ , \ \ \ \
\mathbf{v}_m = V'(1) = l(l+1)+1+m^2 \ee
where we used (\ref{eq20}) and (\ref{eq25r}).

Turning to a calculation of thermodynamic quantities, we start with the free energy which,
at temperature $T$, is
given by the WKB expression~\cite{bibtH}
\be\label{eqFdiv} F = -\frac{1}{\pi} \int_0^\infty \frac{d\o}{e^{\o /T} -1} \int dl (2l+1) p_m(-\infty) \ee
We are interested in the case where $T=T_H$ (eq.~(\ref{eqTH})), but it is convenient to work with
the above ``off-shell'' quantity in order to calculate thermodynamic quantities~\cite{bibsol4}.
The right-hand side of eq.~(\ref{eqFdiv}) is a divergent expression. However, it is not a physical
quantity; the free energy of the system has contributions from the Pauli-Villars fields as well as the gravitational field.
Adding the contributions of the Pauli-Villars regulators, we obtain a regulated expression for the free energy from matter fields
which can be written in terms of
\be p_{\mathrm{reg}} (z) = p_m(z) + 2p_{M_1}(z) - 2p_{M_2} (z) -p_{M_3} (z)\ee
where the various terms are found from (\ref{eq24r}) with the definitions (\ref{eqMMM}).
In the limit $z\to -\infty$, we obtain a finite expression
\be p_{\mathrm{reg}} (-\infty) = \theta_m (\o) + 2\theta_{M_1} (\o)- 2\theta_{M_2}(\o) - \theta_{M_3} (\o)\ee
The free energy~(\ref{eqFdiv}) is corrected to the finite quantity
\be\label{eq32} F_{\mathrm{reg}} \equiv -\frac{1}{\pi} \int_0^\infty \frac{d\o}{e^{\o /T} -1} \int dl (2l+1) p_{\mathrm{reg}}(-\infty) \ee
Using (\ref{eqtheta}), we obtain
\be\label{eq32a} F_{\mathrm{reg}} = - \frac{\pi}{6} A_1T^2 - \frac{2\pi^3}{15} A_2 T^4 + \dots \ee
where we have only exhibited the divergent terms as the Pauli-Villars regulator is removed ($M\to\infty$).

The entropy contribution of the matter field is
\be\label{eq33} S_{\mathrm{reg}} = - \left. \frac{\p F_{\mathrm{reg}}}{\p T} \right|_{T=T_H} =\frac{1}{12} A_1 + \frac{1}{60} A_2 \ee
Including the gravitational contribution~(\ref{eqE}) and (\ref{eqE2}), the total entropy is
\be\label{eq34} S_{\rm{total}} = S_{BH} + \delta S_{gr} + S_{\mathrm{reg}} = \frac{\mathcal{A}_h}{4G_R} - 8\pi\kappa_{2R} + 16\pi \kappa_{3R} \ee
a finite quantity once expressed in terms of the physical constants $G_R$
(eq.~(\ref{eqGren}))
and $\kappa_{2R}$, $\kappa_{3R}$ (\ref{eqk}).


Next, we calculate the two-point function.
We shall calculate the correlator of the time derivative of the scalar
field, $\dot\phi$, rather than of the field $\phi$ itself, in order to avoid
unnecessary complications due to the logarithmic behavior of the (effectively two-dimensional) propagator.

The two-point function at temperature $T_H$ can be written as
\be\label{eq35} G(t,z;t',z') \equiv \langle \{ \dot\phi (t,z,\Omega_2)\,,\, \dot\phi (t',z',\Omega_2')\} \rangle_{T_H}
= \sum_n G_0 (t+in/T_H, z,\Omega_2; t', z',\Omega_2')\ee
in terms of zero-temperature correlators.
The latter may be expanded in a multipole expansion. For a given quantum number $l$,
we obtain the zero temperature two-point function
\be\label{eqgfl} G_0^l (t, z; t', z') = \int_0^\infty d\o\, \o\ e^{-i\o (t-t')} \Phi_{\o l} (z) \Phi_{\o l}^* (z') \ee
Let us set $z=z'$ and $t'=0$ in order to calculate the time correlator.
Moreover, to arrive at explicit expressions, we shall concentrate in the region near the horizon.
Using the approximation~(\ref{eqwzWr}) near the horizon, we obtain
\be\label{eqgfl1} \mathcal{G}_0^l (t) \equiv G_0^l(t,z;0,z) \approx \frac{1}{\pi t^2}  \ee
The $z$-dependent terms are easily seen to give vanishing contributions at the horizon ($z\to-\infty$).
After performing the sum in~(\ref{eq35}), we find
\be\label{eqgfl2} \mathcal{G}^l (t) = \sum_n \mathcal{G}_0^l (t +in/T_H) \approx \frac{\pi T_H^2}{\sinh^2 \pi T_H t}  \ee
Evidently, the two-point function decays exponentially as $t\to \infty$,
\be\label{eq39} \mathcal{G}^l (t) \approx 4\pi T_H^2 e^{- 2\pi T_H t} \ee
independently of the quantum number $l$.

We thus obtained a behavior akin to systems in thermodynamic equilibrium.
However, the entropy of the black hole is large but {\em finite} (eq.~(\ref{eq34})),
including the effects of the matter field.
On the other hand, in the calculation of the two-point function there was no contribution
from the gravitational field.
Its role was to provide the background (\ref{eqmetric}).
In a fixed background, there are no divergences in correlation functions due to curvature,
but the entropy is infinite, as is evident from (\ref{eq33}) if we remove the Pauli-Villars regulator (see eq.~(\ref{eqAAA})).
We need to add the gravitational contribution to make the entropy finite.
Neither the matter nor the gravitational contribution can be independently defined.
If, as expected from the Poincar\'e recurrence theorem, the thermal behavior (\ref{eq39}) ought to receive $\mathcal{O} (e^{-S})$ corrections,
these vanish in a fixed background, because the entropy is infinite.
The corrections are exponentially small but {\em finite} if one includes gravitational effects using the total entropy (\ref{eq34}) instead.
How this comes about is not clear.

In the next section, we assume that gravitational effects amount to a modification of the effective potential (\ref{eq20}) near the horizon and consider various examples.
In all cases, we obtain Poincar\'e recurrences.
The results are independent of the details of the modified potential.

\section{Effective potential}
\label{sec3}

In this section we consider the possibility of quantum effects altering the effective potential (\ref{eq20}) near the horizon.
We consider three explicit examples (a ``brick wall'' \cite{bibtH}, a smooth cutoff and a ``wormhole'' \cite{bibsol})
and show that they all lead to Poincar\'e recurrencies.
Our results at leading order are independent of the shape of the modified potential.

\subsection{A ``brick wall''}

Following 't Hooft \cite{bibtH}, let us
place a brick wall right outside the horizon at
\be\label{eqbw} r=1+h \ \ , \ \ \ \ 0< h\ll 1 \ee
imposing the boundary condition
\be\label{eqbw1} \phi\Big|_{r=1+h} = 0 \ee
In the semi-classical approximation, this boundary condition
restricts the WKB wavefunctions (\ref{eq24r}) to those satisfying the quantization condition
\be\label{eqqc} p_m (z_{\mathrm{min}}) = n\pi - \frac{\pi}{4} \ \ , \ \ \ \
z_{\mathrm{min}} = h+\ln h\approx \ln h \ee
For an explicit expression, we note that we may use the expansion (\ref{eq25r}) for $z\to -\infty$
with a slight (but important) modification,
because we restrict attention to $z\ge z_{\mathrm{min}}$.
We obtain
\be\label{eq41} p_m(z) = - kz + \tilde\theta_m(k) + \dots \ \ , \ \ \ \
\tilde\theta_m(k) = kz_0 + \int_{z_{\mathrm{min}}}^{z_0} \left( \sqrt{\omega^2 - V[r(z')]} - k \right) \ee
where
\be k^2 = \omega^2 - V[r(z_{\mathrm{min}})] \ee
Thus, by inserting the brick wall, we have modified the dispersion relation (\ref{eqdis0}) to one with a small effective mass which is related to the minimum of the potential,
\be\label{eq53r} \omega^2 = k^2 + m_{\mathrm{eff}}^2 \ \ , \ \ \ \ m_{\mathrm{eff}}^2 = V[r(z_{\mathrm{min}})] \approx \mathbf{v}_m h \ee
where $\mathbf{v}_m$ is defined in (\ref{eqtheta}).

The quantization condition (\ref{eqqc}) yields the discrete spectrum of wavenumbers
\be k \approx \left( n + \frac{1}{4} \right) \frac{\pi}{\ln \frac{1}{h}} \ \ , \ \ \ \ n\in\mathbb{Z} \ee
It follows that Green functions are periodic under the radial shift
\be z \to z + L_{\mathrm{eff}} \ \ , \ \ \ \ L_{\mathrm{eff}} = \ln \frac{1}{h} \ee
In particular, using the method of images, the zero temperature two-point function for the $l$th partial wave (\ref{eqgfl})
is modified to
\be\label{eqgflbw} G_0^l (t, z; t', z') = \sum_{n} \int_0^\infty dk\, k\ e^{-i\o (t-t')} \Phi_{\o l} (z+ n L_{\mathrm{eff}}) \Phi_{\o l}^* (z') \ee
We shall calculate this correlator following a similar calculation in~\cite{bibsol2}.
Each term in the series can be written in terms of the two-point function of
a free massive field of mass $m_{\mathrm{eff}}$ in two-dimensional flat space on account of the dispersion relation~(\ref{eq53r}).
Setting $z'=z$ and $t'=0$ for simplicity, we obtain
\be\label{eq47} \mathcal{G}_0^l (t) \equiv G_0^l (t,z; 0,z) = -\frac{1}{2} \sum_{n} \ddot H_0^{(2)} (m_{\mathrm{eff}}\sqrt{t^2 + n^2 L_{\mathrm{eff}}^2}) \ee
For small $t\lesssim 1$, only the term with $n = 0$ contributes (in the other terms, the argument of the Hankel function is approximately
constant, so the time derivative is numerically negligible).
We obtain
\be\label{eq68r} \mathcal{G}_0^l (t) \approx \frac{1}{\pi t^2} \ee
exhibiting a power law decaying behavior for large $t$, which turns into an
exponential decay (eq.~(\ref{eq39})) once temperature effects are included.
This is in agreement with our earlier results without the brick wall (eqs.~(\ref{eqgfl1}) and (\ref{eqgfl2})).

For large $t$ ($t\gtrsim 1$), however, this approximation is no longer valid. For $t\gg \frac{1}{m_{\mathrm{eff}}}$, we
may approximate the sum in eq.~(\ref{eq47}) by an integral.
After some algebra, we arrive at~\cite{bibsol2}
\be\label{eq69r} \mathcal{G}_0^l (t) \approx \frac{\pi m_{\mathrm{eff}}}{2L_{\mathrm{eff}}} e^{-im_{\mathrm{eff}}t} \ee
exhibiting periodicity with period (Poincar\'e time)
\be\label{eq70r} t_P = \frac{2\pi}{m_{\mathrm{eff}}}\ee
Including temperature effects does not alter the above result of periodicity
because the Green function at finite temperature may be written as a series (eq.~(\ref{eq35})) each term of which is periodic with period given by eq.~(\ref{eq70r}).

Expressing the Poincar\'e time in terms of the entropy of the system (eq.~(\ref{eqtP})), we deduce
\be m_{\mathrm{eff}} \sim e^{-S} \ee
Thus the effective mass is exponentially small and may only arise due to non-perturbative effacts.
In terms of the distance parameter $h$, we deduce from (\ref{eq53r})
\be\label{eq52} \sqrt{h} \sim \frac{1}{t_P} \sim e^{-S} \ee
Therefore, the two-point function (\ref{eq69r}) may be written asymptotically in terms of the Poincar\'e time as
\be\label{eq52a} \mathcal{G}_0^l (t) \sim \frac{1}{t_P\ln t_P} e^{-2\pi i t/t_P} \ee
Summarizing, by introducing the brick wall (\ref{eqbw1}), we obtained a modified two-point function which coincided with the one we obtained in the previous section without a brick wall for times $t\lesssim 1 \sim \frac{1}{T_H}$ and exhibited Poincar\'e recurrences for long times $t\gg \frac{1}{T_H}$.

We should point out that with the cutoff (\ref{eqbw1}), we obtain finite thermodynamic quantities.
Indeed, it follows from (\ref{eqFdiv}) that the free energy is finite since $p_m(z)$ should be evaluated at the wall (\ref{eqqc}) instead of at the horizon ($z\to -\infty$).
This yields an expression dependent on $h$ which diverges as $h\to 0$.
Likewise, we obtain an entropy contribution which also depends on $h$ \cite{bibtH}.
However, we may draw no conclusions on the parameter $h$ from entropy considerations, because the total entropy includes a contribution from the Pauli-Villars fields.
The terms that depend on $h$ all cancel and the final expression is independent of $h$ (eq.~(\ref{eq33})).

Next, we replace the sharp cutoff (\ref{eqbw1}) with a smooth modification of the potential (\ref{eq20}) and show that the results remain unchanged at leading order.

\subsection{A smooth cutoff}

Instead of the brick wall, let us make a smooth modification of the potential (\ref{eq20}),
\be\label{eq20sm} V(r) \to V(r) + \frac{\xi^2\mathbf{v}_m}{f(r)} \ \ , \ \ \ \ 0<\xi \ll 1 \ee
where $\mathbf{v}_m$ is defined in (\ref{eqtheta}).
The potential now diverges at the horizon. The change is only significant near the horizon, for $r-1\lesssim \xi $.
To see this, use $V(r)\approx (r-1) \mathbf{v}_m$.
We have $V(r) \sim \frac{\xi^2\mathbf{v}_m}{f(r)}$
when $f(r)\approx r-1\sim \xi $.
The parameter $\xi$ is a physical parameter playing a role similar to the role of $h$ (eq.~(\ref{eqbw})) in the case of a brick wall. We shall calculate its value
by calculating its effects on a physical quantity (the two-point function).

The modification (\ref{eq20sm}) of the potential may be viewed as a change in geometry near the horizon.
Indeed,
suppose that the metric~(\ref{eqmetric}) {\em outside} the horizon changes to
\be\label{eqmetriws} ds^2 = g_{tt} (r) dt^2 + \frac{dr^2}{f(r)} +r^2 d\Omega_2^2 \ \ , \ \ \ \ g_{tt} = - f(r) + \mathcal{O} (\xi^2) \ee
The scalar wave eq.~(\ref{eqwt0}) for the field~(\ref{eqscf}) changes to
\be\label{eqscwws} -\frac{1}{r} \sqrt{-\frac{f(r)}{g_{tt}(r)}}\ \left( r^2\sqrt{-f(r)g_{tt}(r)}\ \left( \frac{\Phi_{\o l}}{r} \right)' \right)' + \frac{\o^2}{-g_{tt}(r)}
\ \Phi_{\o l} - \frac{l(l+1)}{r^2}\ \Phi_{\o l} = m^2 \Phi_{\o l} \ee
To turn this into a Schr\"odinger-like equation, we introduce
the modified ``tortoise coordinate'' $\tilde z$, where
\be\label{eqmtor} \frac{dr}{d\tilde z} = \sqrt{-f(r)g_{tt}(r)} \ee
After some massaging, we obtain from (\ref{eqscwws})
\be\label{eqwtws} -\Phi_{\o l}'' + \{ \tilde V[r(\tilde z)] - \omega^2 \} \Phi_{\o l} = 0 \ee
where a prime now denotes differentiation with respect to $\tilde z$ (\ref{eqmtor})
and the potential is
\be\label{eq20m} \tilde V (r) = -g_{tt}(r) \left[ \frac{l(l+1)}{r^2} +\frac{1}{2r^3} + m^2 \right]
-\frac{f(r)}{2r}\ g_{tt}'(r)
 \ee
replacing eq.~(\ref{eq20}).

Setting
\be\label{eq60} g_{tt} (r) = - f(r) - \frac{\xi^2}{ f(r)} \ee
the modified potential (\ref{eq20m}) becomes
\be\label{eq20V} \tilde V (r) = V(r) + \delta V \ \ , \ \ \ \
\delta V = \frac{\xi^2}{ f(r)} \left[ \frac{l(l+1)}{r^2} + m^2 \right] \ee
Near the horizon, the additional contribution is $\delta V \approx \xi^2 (\mathbf{v}_m -1)/f(r)$,
matching the desired form (\ref{eq20sm}) with the substitution $\mathbf{v}_m \to \mathbf{v}_m -1$.
Away from the horizon, the modification is negligible.

Next, we proceed to solve the Schr\"odinger-like wave equation (\ref{eqwtws}).
The modified ``tortoise coordinate'' is found explicitly as a function of $r$ by integrating (\ref{eqmtor}),
\be \tilde z = \frac{r\sqrt{f^2(r) +\xi^2}}{1+\xi^2} + \frac{1}{(1+\xi^2)^{3/2}}
\ln \frac{r}{2} \left[ \frac{f(r)+\xi^2}{\sqrt{1+\xi^2}} + \sqrt{f^2(r)+\xi^2} \right] \ee
The horizon ($r=1$) is mapped at
\be\label{eq63} \tilde z_h = \frac{\xi}{1+\xi^2} + \frac{1}{(1+\xi^2)^{3/2}}
\ln \frac{1}{2} \left[ \frac{\xi^2}{\sqrt{1+\xi^2}} + \xi \right] \approx \ln \frac{\xi}{2} \ee
We therefore need to solve the Schr\"odinger equation in the interval $[z_h,\infty)$.
The minimum of the potential (\ref{eq20V}) is at
\be r_{\mathrm{min}} \approx 1+ \xi\sqrt{\frac{\mathbf{v}_m - 1}{\mathbf{v}_m}} \ee
which in terms of $\tilde z$ is very close to $\tilde z_h$.
As before ({\em cf.}~eq.~(\ref{eq53r})), the minimum value of the potential provides an effective mass,
\be\label{eq65} m_{\mathrm{eff}}^2 = \tilde V(r_{\mathrm{min}}) \approx 2\xi\sqrt{\mathbf{v}_m (\mathbf{v}_m -1)} \ee
In the WKB approximation the wavefunction for $\tilde z\in [\tilde z_1, \tilde z_2]$ reads
\be\label{eq24s} \Phi_{\o l}^{\mathrm{WKB}} (\tilde z) = \mathcal{C} \sin \left( \tilde p_m (\tilde z) + \frac{\pi}{4} \right)
\ \ , \ \ \ \ \tilde p_m (\tilde z) = \int_{\tilde z}^{\tilde z_2} d\tilde z' \sqrt{\o^2 - \tilde V[ r(\tilde z')]} \ee
where $\tilde z_{1,2}$ are zeroes of the expression under the square root.
Frequencies are quantized under the Bohr-Sommerfeld quantization condition
\be\label{eq52s} \tilde p_m (\tilde z_1) = n\pi + \frac{\pi}{2} \ee
The larger turning point $\tilde z_2$ is in a region where the modification of the potential is negligible,
therefore $\tilde z_2 \approx z_0$ (see eq.~(\ref{eq24r})).
The smaller turning point $\tilde z_1$ is even closer to the horizon than the minimum of the potential, therefore
$\tilde z_1\approx \tilde z_h$ (eq.~(\ref{eq63})). Expanding around the minimum of the potential, we obtain from (\ref{eq24s}),
\be \tilde p_m (\tilde z) \approx -k\tilde z \ \ , \ \ \ \ \omega^2 = k^2+ m_{\mathrm{eff}}^2 \ee
The quantization condition (\ref{eq52s}) then yields
\be\label{eq65a} k \approx \left( n + \frac{1}{2} \right) \frac{\pi}{L_{\mathrm{eff}}}
\ \ , \ \ \ \ L_{\mathrm{eff}} = \ln \frac{2}{\xi} \ee
where we used (\ref{eq63}).

The calculation of the two-point function proceeds as in the case of the brick wall. We obtain the decaying behavior of eq.~(\ref{eq68r}) for times $t\lesssim \frac{1}{T_H}$.
For longer times ($t\gg \frac{1}{T_H}$), we obtain the oscillatory behavior of
eq.~(\ref{eq69r}).
Using (\ref{eq70r}) and (\ref{eq65}), we obtain the parameter $\xi$,
\be\label{eqxi} \sqrt{\xi} \sim \frac{1}{t_P} \sim e^{-S} \ee
as with $h$ in the case of a brick wall (eq.~(\ref{eq52})).

Substituting the values of the parameters (\ref{eq65}), (\ref{eq65a}) and (\ref{eqxi})
in eq.~(\ref{eq69r}), we obtain the asymptotic form of the two-point function
\be\label{eq52b} \mathcal{G}_0^l (t) \sim \frac{1}{t_P\ln t_P} e^{-2\pi i t/t_P} \ee
which coincides with our earlier result (\ref{eq52a}) for the brick wall.

\subsection{A ``wormhole''}

The smooth cutoff we considered above amounted to a modification of the geometry near the horizon which was singular at the horizon (eq.~(\ref{eq60})).
Here, we consider a case in which $g_{tt}$ approaches a (non-vanishing) finite value at the horizon.
This modification effectively replaces the horizon with the throat of a wormhole~\cite{bibsol}.

For explicit calculations, we set
\be\label{eq42} g_{tt} (r) = -f(r)-\lambda^2 \ \ , \ \ \ \ 0 < \lambda \ll 1 \ee
This modification leads to a modified ``tortoise coordinate'' $\tilde z$ (\ref{eqmtor}) and a Schr\"odinger-like wave equation (\ref{eqwtws}) with the modified potential (\ref{eq20m}).
By integrating (\ref{eqmtor}), we obtain
\bea \tilde z &=& \frac{\sqrt{(r-1)[(1+\lambda^2)r -1]}}{1+\lambda^2} \nonumber\\
&+&
\frac{1+\lambda^2/2}{(1+\lambda^2)^{3/2}} \ln \frac{\sqrt{1+\lambda^2}}{\lambda^2/2} \left[ \frac{(1+\lambda^2)r-1-\lambda^2/2}{\sqrt{1+\lambda^2}} + \sqrt{(r-1)[(1+\lambda^2)r -1]} \right]
\label{eq73}\eea
where we chose the integration constant so that $\tilde z = 0$ at the horizon ($r=1$) for convenience.
Thus, the region outside the horizon is mapped onto the positive real axis on the $\tilde z$ plane.
However, we are not to solve the Schr\"odinger equation (\ref{eqwtws}) for $\tilde z \in [0,\infty)$,
because the modified potential is regular at $\tilde z = 0$.
Indeed, from (\ref{eq20m}) and (\ref{eq42}), we obtain
\be\label{eqwtw} \tilde V(r) = V(r) + \delta V \ \ , \ \ \ \
\delta V = \lambda^2 \left[ \frac{l(l+1)}{r^2} +\frac{1}{2r^3} +m^2 \right]
 \ee
Unlike the previous case (eq.~(\ref{eq20V})), $\delta V$ is regular and has a finite value at the horizon.
Therefore, there is no obvious boundary condition at what was the horizon before the modification (\ref{eq42}) that one can impose.
In fact, in the absence of a horizon, there is no reason not to include the region $\tilde z < 0$.

To understand the geometric meaning of this continuation, notice that $\tilde z$ is only defined for $r>1$.
The unobservable region inside the horizon of the black hole is inaccessible after the modification (\ref{eq42}). This is because only $g_{rr} = \frac{1}{f(r)}$ (eq.~(\ref{eqmetric})) changes sign if we cross the horizon and {\em not} $g_{tt}$ (eq.~(\ref{eq42})).
The modified metric in terms of $\tilde z$ reads
\be\label{eqmetricm} ds^2 = \left( 1- \frac{1}{r} + \lambda^2 \right) [ -dt^2 + d\tilde z^2] + r^2 d\Omega_2^2 \ee
where we used (\ref{eqmtor}) and (\ref{eq42}).
The coordinate $r$ is given in terms of $\tilde z$ through (\ref{eq73}) for $r>1$.
Thus (\ref{eqmetricm}) is the metric for the patch $\tilde z >0$.
Picking the negative square root in (\ref{eqmtor}) leads to the same metric defined on the patch $\tilde z <0$.
In the latter patch, $r$ is still given by (\ref{eq73}) but with $\tilde z$ replaced by $-\tilde z$.
The two patches are bridged together by a {\em wormhole} at $\tilde z = 0$.
The resulting metric is smooth. This can be seen by zooming onto the throat at $\tilde z = 0$.
Eq.~(\ref{eq73}) may then be inverted to give
\be r-1 \approx f(r) \approx \lambda^2 \sinh^2 \frac{\tilde z}{2} \ee
and the metric (\ref{eqmetricm}) becomes
\be ds^2 \approx \lambda^2 \cosh^2 \frac{\tilde z}{2} [ -dt^2 + d\tilde z^2] + d\Omega_2^2 \ee
with no singularity at $\tilde z = 0$.

It follows that we ought to solve the wave equation in the entire spacetime defined by (\ref{eqmetricm}) which includes both sides of the wormhole.
Therefore, we shall solve the Schr\"odinger equation (\ref{eqwtws}) for $\tilde z \in \mathbb{R}$.
The modified potential (\ref{eqwtw}) is an even function of $\tilde z$.
Its minimum is at $\tilde z = 0$. As before, it provides an effective mass
\be\label{eq65w} m_{\mathrm{eff}}^2 = \tilde V (1) = \lambda^2 \left[ \mathbf{v}_m - \frac{1}{2} \right] \ee
to be compared with (\ref{eq53r}) and (\ref{eq65}).

In the WKB approximation, we obtain the wavefunctions
\be\label{eq24sa} \Phi_{\o l}^{\mathrm{WKB}} (\tilde z) = \mathcal{C} \sin \left( \tilde p_m (\tilde z) + \frac{\pi}{4} \right)
\ \ , \ \ \ \ \tilde p_m (\tilde z) = \int_{\tilde z}^{\tilde z_0} d\tilde z' \sqrt{\o^2 - \tilde V[ r(\tilde z')]} \ee
for $|\tilde z| \le \tilde z_0$, where $\tilde V[ r(\tilde z_0)] = \o^2$.
The explicit form of $\tilde z_0$ is found by comparing with the wave equation discussed in section \ref{sec2}. Notice that the modified tortoise coordinate $\tilde z$ (\ref{eq73}) is related to the tortoise coordinate (\ref{eqtor}) away from the horizon (for $r-1 \gg \lambda$) by
\be\label{eqttor} \tilde z \approx \ln \frac{2}{\lambda^2} + z \ee
It follows from (\ref{eq24r}) that
\be\label{eqttor1} \tilde z_0 \approx \ln \frac{2}{\lambda^2} + z_0 \ee
The explicit form of $z_0$ is not needed, because $z_0 \ll \ln \frac{2}{\lambda^2}$.

The Bohr-Sommerfeld quantization condition reads
\be\label{eq52w} \tilde p_m (-\tilde z_0) = 2p_m (0) = n\pi + \frac{\pi}{2} \ee
Using (\ref{eq41}), after substituting the tortoise coordinate and potential of the wormhole, we deduce
\be\label{eq41w} \tilde p_m (\tilde z) \approx -k(\tilde z -\tilde z_0) \ \ , \ \ \ \ \omega^2 = k^2 + m_{\mathrm{eff}}^2 \ee
The quantization condition (\ref{eq52w}) reads
\be\label{eq52w1} k \approx \left( n + \frac{1}{2} \right) \frac{\pi}{L_{\mathrm{eff}}} \ \ , \ \ \ \ L_{\mathrm{eff}} = 2\ln \frac{2}{\lambda^2} \ee
We may now calculate the two-point function as in the other two cases.
We obtain the decaying behavior of eq.~(\ref{eq68r}) for times $t\lesssim \frac{1}{T_H}$, whereas for times $t\gg \frac{1}{T_H}$ we obtain the oscillatory behavior of eq.~(\ref{eq69r}).

The parameter $\lambda$ may then be deduced from (\ref{eq70r}) and (\ref{eq65w}),
\be\label{eqlam} \lambda \sim \frac{1}{t_P} \sim e^{-S} \ee
The asymptotic form of the two-point function (\ref{eq69r}) may be written in terms of the Poincar\'e time using (\ref{eq65w}), (\ref{eq52w1}) and (\ref{eqlam}),
\be\label{eq52c} \mathcal{G}_0^l (t) \sim \frac{1}{t_P\ln t_P} e^{-2\pi i t/t_P} \ee
Thus, even though the geometry was different, the asymptotic form of the time correlator proved to be the same as in the other two cases (eqs.~(\ref{eq52a}) and (\ref{eq52b}), respectively).

\section{Conclusions}
\label{sec4}

We discussed the dynamics of a massive scalar field outside the horizon of a Schwarzschild black hole.
In the semi-classical approximation, one obtains time correlators which decay exponentially
as expected of correlators in a thermal bath.
Since the black hole has finite entropy $S$, if one waits long enough, one should observe Poincar\'e recurrences with long period (Poincar\'e time) $t_P \sim e^S$.
The thermal correlators should contain corrections $\mathcal{O} (e^{-S})$ which will become important at later times thus restoring unitarity.
Understanding how this emerges in the dynamics of a matter field is not straightforward because the latter contributes an infinite amount to the
entropy of the black hole.
Consequently, the $\mathcal{O} (e^{-S})$ corrections one expects vanish.
The infinite corrections to the entropy were understood in terms of standard renormalization of the gravitational parameters (Newton's constant, etc) \cite{bibE2}.
The $\mathcal{O} (e^{-S})$ corrections to correlators of matter fields are yet to be understood.

Without trying to probe the origin of these corrections, we considered examples of possible modifications of the effective potential of matter fields
due to quantum effects.
Using a semi-classical approximation, we calculated the time correlator of a massive scalar field and obtained the same asymptotic form in all cases we considered (eq.~(\ref{eq52a})) exhibiting Poincar\'e recurrences.
Thus results appear to be independent of the details of the modification of the potential.
These modifications may be attributed to a change in the geometry near the horizon due to quantum effects.
In some cases, the geometrical picture seems to differ drastically.
For example, in the case of a ``wormhole'' \cite{bibsol}, the modification seems to lead to inaccessibility of the interior of the black hole including the singularity (for the geometry with metric (\ref{eqmetricm})).
However, these modifications are not valid away from the horizon at a macroscopic distance where they should be matched with the Schwarzschild metric (\ref{eqmetric}).

All cases we considered shared key features: the potential had a minimum which led to a small but finite effective mass $m_{\mathrm{eff}} \sim 1/t_P \sim e^{-S}$
parametrizing the corrections to the thermal correlators.
Moreover, the potential was fairly flat over a distance $L_{\mathrm{eff}} \sim \ln t_P \sim S$ in the radial direction (in terms of the ``tortoise coordinate'') rising beyond the flat region.
This allowed the analytic solution of the Schr\"odinger-like wave equation in the semi-classical approximation.
Details of the shape were not important at leading order.

It should be pointed out that our analysis concentrated on the region {\em outside} the horizon.
It would be interesting to apply a similar analysis to the unobservable region inside the horizon and match the results across the horizon.
This ought to shed light on the evolution of a black hole beyond the thermal state of Hawking radiation.

\section*{Acknowledgments}

I wish to thank M.~Einhorn and S.~Mathur for discussions.

\end{document}